# Towards Realizing the Smart Product Traceability System

Dharmendra K Mishra, Yacine Ouzrout
University Lyon 2 Lumiere,
DISP Laboratory, France
{dharmendra-kumar.mishra, yacine.ouzrout}@univ-lyon2.fr

[1]Ajay K shrestha, [2]Abdelaziz Bouras
[1]Kantipur Engineering College, Kathmandu
[2]Qatar University, Qatar

*Abstract* :

The rapid technological enhancement and innovations in current days have changed people's thought. The use of Information Technology tools in people's daily life has changed their life style completely. The advent of various innovative smart products in the market has tremendous impact on people's lifestyle. They want to know their heart beat while they run, they need a smart car which makes them alert when they become sleepy while driving, and they need an IT tool which can make their home safer when they are out. These quests of people to the very extent have been fulfilled by development of many Meta products like wearables, smart phones, smart car etc. The concept of Meta product is based on the fact that they need to offer intelligent services to the users. Its intelligence level is higher than the smart and other intelligent products. There are various stakeholders in the Meta products environment viz. manufacturers, users, experts, and third party service developers. A common collaborative platform that can be used by all these stake holders is necessary to innovate the Meta products services. A cloud architecture is used to achieve the innovations. The aim of this paper is to propose a collaborative framework based on cloud to achieve the various IT services in the meta product's environment.

I. INTRODUCTION

Technological Innovations in recent years have led to development of many computing infrastructure such as cloud computing, intelligent computing, RFID, wearable sensor devices etc. The advent of advancement in these technologies has given birth to many products which fulfill the changing needs of the people. Intelligent products, smart products and Meta products are in the market with different level of intelligence. The intelligence level of Meta products is higher than the other intelligent products. These products communicate with environment, have decision making abilities and can recognize themselves through the product identification technologies. .

Meta products are comprised of many other components to offer intelligent services to the users. Apple's iPOD is considered to be first of its kind. As per [1], "Meta Products are dedicated network of services, people, products and environment fed by the information flows made possible by the web and other ubiquitous technologies: web-enabled-product-services networks". There are various types and purposes of Meta Products. The sensors in a smart T-shirt for example records the user's heart beat and stores the data in the cloud for investigation by the experts. Similarly Google's glass is the recent example of Meta Products.

The data generated by the Meta product's services are stored in the cloud from where users can extract the required information. The services offered by the Meta products are needed to be innovative and enhanced. The manufacturers need user's feedback to design innovative product, experts need to answer the user's queries in the required time and they also need to give users the proper advice. Similarly, third party service providers need to develop the service as per the requirement of the product and users must need to be assured that their personal information are secured and they can access it any time they want. To achieve all these requirements a collaborative platform is needed to be developed that can be used by all the stake holders of the Meta product's environment. Cloud Computing is a popular technology of these days that can be used to achieve collaboration. . Cloud computing is a concept of virtualization through which user fulfill their IT demand as a service through internet. It offers the service in its three service model:- (1) Software-as-a-Service(SaaS), (2) Platform-as-a-Service(PaaS) and Infrastructure-as-a-Service(IaaS). With SaaS model, service providers provide users the different application software as a service at a very low cost. Users don't have to buy the software rather they just have to pay for what they use. Developers develop their application by using cloud platform in the PaaS model and using IaaS model, users access software and hardware infrastructure for example they can store their data in the cloud server. The purpose of this research work is to propose a common cloud framework that can be used by all the stakeholders of Meta products environment for the various purposes. For the framework a PaaS architecture is proposed.

The rest of the paper is organized as follow: In section 2 we describe the overview of Meta-Products. Review of existing platforms is presented in section 3 with their limitations. In section 4 we propose our model for the integrated platform and finally section 5 will conclude our work.

## II. META- PRODUCTS OVERVIEW

In today's world we have the technological as well as societal capacity that can revolutionize ours daily life style. The transition we are living now is bringing information-fuelled products and services that will be around us as a network whenever we need them. The result of finding (or designing) these increasingly clever ways to use information is what we call Meta Products [1]. A Meta-product is a customer driven customized entity having computing and sensory units that in turn are connected with cloud to providing the intelligent services to the users.

Meta-Products are defined by [1] as: "Web-enabled product-service networks". Meta-Products consist of physical and web elements. Physical elements consist of sensors which gives input to the web elements where processing is done. Similarly, actuators are also physical elements that are used by web elements to alert or inform the users. The web elements (or services) enable Meta-Products to be multi-functional, efficient, valuable and exciting to users [2]. The overall concept of the Meta Products is explained by the following figure.

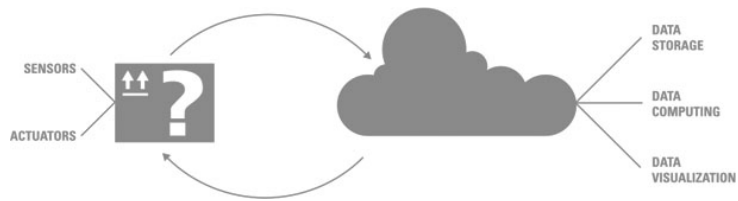

Fig. 1. Meta Products Elements [1]

In the above figure, sensors sense the required data depending on the nature of the product. These data are put into the cloud for storage, visualization and computing. In reality, meta-Products consists in three main parts, physical part (as any traditional product), electronic part (consisting on sensor networks) and web elements (services to operate them). There are different types of Meta products having different purposes. A wearable for example tracks the user's health condition and inform them accordingly after proper analysis. Similarly a smart home system protects itself from fire, theft, and others and informs the owner accordingly.

In a nutshell the services offered by the Meta products can be studied under 4 headings: (1) Monitoring, (2) Control, (3) Optimization and (4) Automation [3]. The sensors and other data sources monitor the product's condition, operation and external environment and inform the users accordingly. The sensors in the car for example monitor the weather condition and inform the driver accordingly. These connected products can be automatically controlled by some software in the cloud. For example, alert the driver when he gets lousy while driving. Smart products use algorithms to analyze and use historic data and optimize it to take innovative decision. A microcontroller in wind turbine for example adjust blade in every revolution to get maximum wind energy [3]. The sensors for monitoring and some algorithms for control and optimization make smart product to act automatically. A vacuum cleaner for example cleans the surface in different orientation; a refrigerator controls the temperature sensing the food inside.

Nowadays, Meta-Product manufacturers propose services specific for one Meta-Product or available only for this manufacturer's products. This method conflicts with the concept of Meta-Products considered as multi-functional and combinable. Other than intelligent or smart products, Meta products have complex organization and offer the services in more intelligent way to its users. Next section will introduce the existing platform and their limitations.

## III. RELATED WORKS

As discussed in above section, there are various industries working on the Meta-Product's designing and manufacturing. All of them have their own platform for the specific product. In this section we describe the current literature on the existing Met-Product's integrated platform.

The term smart product sometimes is used to define Meta products. Smart products use sensors, actuators, microcontroller and internet for communication to offer intelligent services to users.

Kortuem G., Kawsar F et al. in their work developed a smart object deployed at a road construction site. Workers wear this device that showed personal health records containing information about their exposure to hazardous equipment vibration. In this work the authors uses wearable sensors to record the data and displayed the information [4].

Kaur P.D. and Chana I., in 2013 have proposed a Cloud Based Intelligent Health Care Service (CBIHCS) that performs real time monitoring of user's health data for diagnosis of chronic illness such as diabetes. In their work "Cloud based intelligent system for delivering health care as a service" the authors use different types of sensors to record health information for example blood pressure, glucose level, weight scale etc. and these data are transferred by mobile or PDA into the cloud. These data are analyzed using the prototype developed by authors at SaaS level in the cloud system. This information is transferred to the concerned experts whose suggestions are again stored in the cloud infrastructure. Users can then extract the required information in a secured way through the user sub system. [5].

Wuryandary A., Gondokaryono Y et al. in 2013 developed a driver main computer system which records and gives driver the information of car and environment, it also alerts driver to protect from possible accidents towards having a smart car [6].

Mazzei D., Montelisciani G., et al in 2014 proposed a web platform for the design, co-design and sharing of smart object in their work "Internet of Things for Designing Smart Objects" [7] In this work, platform allows users three virtual areas of the smart product: Body, Brain and Behavior. Through the body, users can select the shape of the product, in the brain area, sensors and other electronic device extract and analyze the data and finally behavior part explains the functionality of the device [7].

In another work, wearables sensors and a cloud based platform to monitor environmental parameters in e-health Applications has been discussed [8]. The wearables in this work records the body temperature which is transferred into the cloud through microprocessor for storage and analysis.

Similarly, Khan F., Ali H. et al in 2014 proposed a Cloud based healthcare framework for security and patient's data privacy [9]. The model works in two fold. First: the indoor, in which the patient is admitted into the hospital. In this case the sensors attached in the patient's body records his physiological data which is the transferred to the hospital cloud. The next on is outdoor in which the patient is not admitted in hospital. In this case patient is connected with remote base station which then transfers the data into the hospital community cloud. The authors developed an algorithm to maintain privacy and security of the patient's data [9].

There exists some platforms based on Internet of Things (IoT) developed by various companies for the smart products. IoT is an environment that allows everything either people, object or other to connect themselves and able to exchange information wirelessly through the smart sensors attached to it. IoT use cloud computing technology to store and compute the data extracted by sensors or other PEID devices.

Everything's IoT smart products platforms provide highly scalable data management and application services for smart connected products. It captures and collates millions of data about the products from their conception to end throughout their life cycle, which are analyzed by the design team to design new innovative products [10]. Similarly, Swarm is an IoT development platform that facilitates adding new services to products easily [11]

OpenRemote is an open-source middleware solution for the Internet of Things. It allows users to integrate any device regardless of brand or protocol and design any user interface for iOS, Android or web browsers [11].

IoBridhge is another platform that allows connecting any product to a mobile device via the web. It reduces the time to market and cost per product helping in proper decision making process [11].

A new cloud-based infrastructure Zatar automatically detects the devices and connects them to the Internet, through which all the connected devices and their respective users can share data and collaborate seamlessly [11].

A cloud-based application enablement platform, Ayla Networks is a simple and cost-effective solution for to connect any device to the Internet. It provides powerful software agents embedded in both connected devices and mobile device applications for end-to-end support [11].

. The recent launch of Google glass is another innovation in the Meta product's Industries. The development in Google glass can be done by two API's (i) Mirror API and (ii) Glass development Kit (GDK). With Mirror API users can create RESTful web services. But the problem with this API is that it cannot work offline. GDK is an add-on to the Android SDK. It allows running application offline also.

. Similarly, Arrayent for example is the best IoT platform supporting millions of devices. It provides M2M platform with lower-level web service API and interfaces to a presentation layer so that unique, branded user interfaces can be developed for the products [12]. However the platform enables only some major brands viz. Whirlpool, Maytag and First Alert to bring smart, connected devices to consumers.

The AllSeen Alliance is another vibrant, collaborative open source effort which has released two versions of AllJoyn including SDKs for developers and appliance manufacturers to easily introduce connected appliances to the market. Commercial products use AllJoyn today, from LG HDTVs to LIFX smart light bulbs [13].

Most importantly there is another Sine-Wave Technologies' open platform with a hosted set of APIs that provides a support required to construct and manage IoT applications [14]. The platform enables a rapid deployment of branded remote asset management solutions for enterprise companies by supporting smart devices from any vendor as well as custom-built devices.

.Considering the life cycle of the Meta product, there is lot of data generated in its use phase. All the stakeholders use these data for different purpose. For example, manufacturers need the user's feedback data to improve the design of the product; users need to know the expert advice and third party need to know what are the requirements of the product. A common and collaborative platform for sharing the data that can fulfill all these requirements is essential to improve the services offered by the Meta product development services, which is missing in the literature.  Maximum of Meta products use SaaS model only for the data storage and processing. All of the above discussed platforms are either beneficial for the product's manufacturer to design new improved products or for the users who can use their limited functionality only. . There is no common platform exist that can be used by users, developers, manufacturers, or third party service providers in the use phase of product's life cycle. The enterprises associated with manufacturing of such type of product must need to act in collaboration to achieve the above requirements. To overcome this limitation, we propose a common platform based on cloud that can be used to get different services like data security, processing, getting expert advice and so on, in the following section. The concept of Cloud Computing helps to develop a collaborative platform that can be used by every stakeholders to achieve their tasks in an efficient way.

## IV. PROPOSED PLATFORM

It fulfils every demand of the customers, as a service through internet. It allows customers to pay for the services they have subscribed. They don't need to pay for the infrastructure. The entire infrastructure is managed by the cloud service providers. For this reason, Cloud Computing is also called as "Pay-as-you-Go" model just like the public utility services.

The vendor in charge of the platform holds a set of software and development tools in its server and the developers use their API (Application Program Interface) to develop their own application at the remote server, supporting the complete life cycle of building and delivering web applications.

Using the SaaS model, the platform will offer various Meta-Products services to their customers. The various services are hosted in platform by the vendor and made available to customers over a network, typically the Internet allowing them to choose as per their needs.

The proposed integrated platform will scale for millions of connected devices with no degradation of performance, reliability or security. All of the capabilities needed to rapidly provide highly scalable data management and application services that includes the construction, integration, diffusion and monetization of composite services across connected Meta-Products (e.g. wearables, large and small household appliances, consumer electronics, various sensors, PCs, Tablets and so on) are gathered together in one common integrated platform as given in Figure 2.

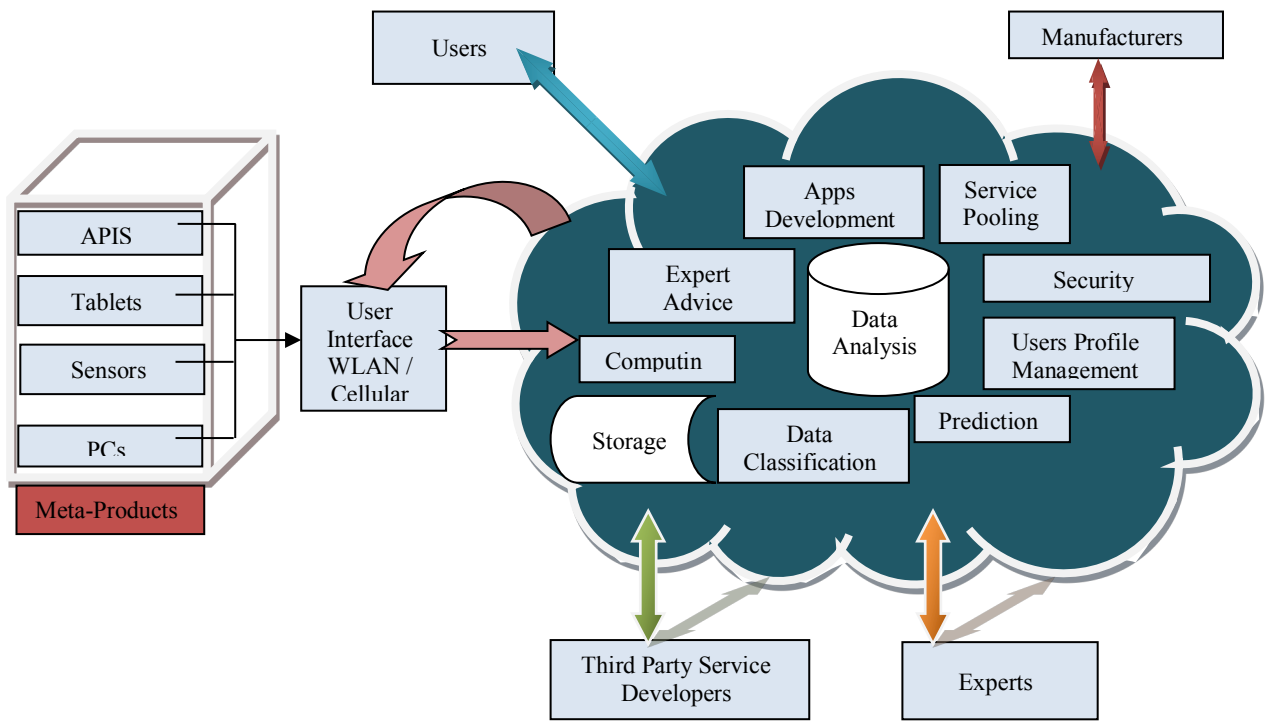

Fig. 2 Proposed Architecture of Integrated Cloud based Platform

The proposed platform offers almost all the IT needs as the services. The services available are described in the following paragraph.

- *Computing:* With this service, the data sent by the connected products are processed at the server and made available for the users.
- *Storage:* There is unlimited data storage space available with cloud. Users can easily access the stored data from anywhere.
- *Security:* As data from the Meta Products are very sensitive, the control of trust in the cloud is one of the most significant aspects. Slight tampering on the Meta Products' data may lead to very dreadful situation. So the proposed platform must be also designed carefully so as to provide efficient data security solutions and services to protect user's data in the cloud. Besides, there still exists a crucial problem with transforming human behaviour into numbers for performing the measurement.
- *Data Analysis:* With this service, the platform analyses the data sent by the connected products comparing with other data and give users the predicted result. Sensor's data of the heart beat for example is analyzed with some historic data.
- *Prediction:* Based on the analysed data discussed above, users can see the predicted result for example health condition, weather condition etc.
- *Expert Advice:* With this service, users can have the different expert advices. The data sent by the wearable about blood pressure and heart beat while running for example, are analyzed by the doctors who will give their advices to the users through the platform.
- *Users profile management:* With this service, the platform manages the history and data of the user's profile. The products manufacturer can have idea about how many users are using their products. They can see the user's feedback that will be helpful for them while designing the new product.
- *Data Classification:* There are various types of data sent by the connected products. Some data may be related with health, while some may be related with weather condition or about the products performance. All these different types of data are needed by the different users. With this service the data are classified as per their type and made available for the users who can easily access the required data.
- *Apps Development:* With the PaaS model of the platform the developers can develop their own application for the various Meta products. For this they don't have to buy the platforms, they just need to pay for the instances they use.

The platform can be exploited and generate value in multiple levels for various stakeholders:

- Manufacturers to enable their product on the Cloud. They could be charged when adding new Meta-Products to platform.
- Users to benefit from the services. They can be charged first when registering their Meta-Products on the platform. Then, charging could be also per service for instance when using the storage service, the user could pay according to the usage range.
- Third party developers that can develop new services and offer them or sell them through the services store. The percentage of the revenue of sold services could go to the platform in return for the management, maintenance, tools for development and other services available for the developers.
- Experts that can requested to analyse collected information to inform users and advise them about their current health status for example. Experts are also charging for the services they are proposing and part of this could be shared with platform.

## V. CONCLUSION

There are various IoT platforms exist today developed by different developers. Purpose of all these platforms is to provide developers and users to add some services, access the data, giving alert etc with the many products connected with these platforms. Some platforms are developed for some specific products only. There exists no any platform that can be used by all the interested stakeholders of the Meta products environment. Thus, in this work, we purpose to develop a common integrated cloud based platform that can be used by any Meta product users, developers or others for their own purposes. The services offered by the proposed platform will satisfy required needs of all the stakeholders.

**ACKNOWLEDGEMENTS**

The authors would like to thank the European Commission for their financial support (Easy-Imp project grant agreement no 609078) and Erasmus Mundus cLINK project.